\documentclass[a4paper,12pt]{article}

\usepackage[english]{babel}

\usepackage[linktocpage,unicode]{hyperref}
\usepackage{csquotes,graphicx,fancyhdr}
\usepackage{setspace,longtable,graphicx,hyphenat,hyperref,fancyhdr,ifthen,enumitem,amsmath}
\hypersetup{unicode=true,colorlinks=true,linkcolor=black,citecolor=blue,urlcolor=blue}
\usepackage{url}

\usepackage{geometry}
\usepackage{amsfonts}%
\usepackage{amsthm}%
\usepackage{mathrsfs}%
\usepackage[title]{appendix}%

\usepackage{textcomp}%
\usepackage{manyfoot}%
\usepackage{algorithm}%
\usepackage{algorithmicx}%
\usepackage{algpseudocode}%
\usepackage{listings}

\newcommand{\e}{\text{e}}

\newcommand{\D}{\text{D}}
\newcommand{\M}{\text{M}}
\newcommand{\N}{\text{N}}
\newcommand{\K}{\text{K}}
\newcommand{\CC}{\text{C}}

\newcommand{\LL}{\text{L}}
\newcommand{\n}{\text{n}}

\def\beq{\begin{equation}}
\def\eeq{\end{equation}}
\def\bear{\begin{eqnarray}}
\def\ear{\end{eqnarray}}
\def\rf{\eqref}
\def\nn{\nonumber\\ {}}

\def\red#1{{ #1}}

\def\email#1#2{\footnotetext[#1]{e-mail: #2}\addtocounter{footnote}{1}}
\newcommand{\Pl}{\text{Pl}}
\newcommand{\HH}{\text{H}}

\title{\bf Spontaneous Brane Formation}

\author{{Arkadiy A. Popov} $^{a,1}$, Sergey G. Rubin $^{a,b,2}$}

\date{\small\it
$^{a}$ \quad N.~I.~Lobachevsky Institute of Mathematics and Mechanics, {Kazan  Federal  University,}
 Kremlyovskaya 18, Kazan 420008, Russia\\
$^{b}$ \quad Moscow Engineering Physics Institute, National Research Nuclear University MEPhI,  Kashirskoe shosse 31, Moscow 115409, {Russia}}

\begin{document}
\maketitle
\email{1}{apopov@kpfu.ru}
\email{2}{sergeirubin@list.ru}

\abstract{This paper presents a study of brane formation in six-dimensional space. There is no a priori assumption of the existence of brane(s). However, an analysis of the generalized Einstein equations shows that there { is} a set of metrics describing two static branes even in the absence of matter fields. At the same time, no one-brane configurations were found. The trapping of massive particles on branes is a consequence of the metric structure, which prevents these particles from moving between branes.
It is shown that communication between charged particles on different branes is provided by photons. Such positron--electron annihilation could be studied experimentally at the LHC collider.
The Higgs field is distributed between the branes in such a way that it can serve as a Higgs portal connecting two worlds located on different branes. The values of the 4D physical parameters depend on the extra metric structure near the branes.
We also found a non-trivial effect of the decompactification of extra space during the Hubble parameter variation.
}


\section{Introduction}\label{sec1}
The idea of multidimensional gravity is an important tool for obtaining new theoretical results~\cite{Abbott:1984ba,Chaichian:2000az,Brown:2013fba,Bronnikov:2009zza}.
In paper~\cite{Krause:2000uj}, warped geometry is used to solve the problem of the small cosmological constant. The concept of multidimensional inflation is discussed in detail in several sources, including~\cite{2002PhRvD..65j5022G,Bronnikov:2009ai,Fabris:2019ecx}. These papers postulate that an extra-dimensional metric $g_n$ is stabilized at a high energy scale. The stabilization of extra space as a purely gravitational effect has been studied in~\cite{2003PhRvD..68d4010G,Arbuzov:2021yai}.

Theories concerning fields confined to a three-dimensional hypersurface (a brane) within a multidimensional spacetime have been a significant area of interest for several decades. Akama was the first to propose this idea~\cite{Akama:1982jy}. The thick brane hypothesis was independently proposed by Rubakov and Shaposhnikov~\cite{Rubakov:1983bb}.
Branes are an essential tool for solving a wide range of problems. The hierarchy problem is undoubtedly one of the most important such problems~\cite{Gogberashvili:1998vx,1999PhRvL..83.3370R,ArkaniHamed:1998rs}.

Thin branes were the first to be discussed in the literature. It has been shown that they suffer from some intrinsic problems~\cite{Burgess:2015kda}. So, despite hopeful results, they are gradually being replaced by branes with an internal structure---thick branes~\cite{Bronnikov:2006bu,Chumbes:2011zt,Hashemi_2018,Dzhunushaliev:2019wvv,Bazeia:2022vac,Wan:2020smy}. The latter require special conditions to prove their existence and stability. It could be the choice of a specific metric, including the warp factor~\cite{Oda_2000} or  a special form of the scalar field potential~\cite{Bazeia:2022vac}. A pure geometric approach based on $f(R)$ gravity in 5D is performed in~\cite{Guo:2023mki} where a substantial number of bibliographic sources can be found.

In brane world scenarios, the key issue is the localization of fields on branes and the recovery of effective 4D gravity.
The localization of gauge and spinor fields on branes is a necessary element of any brane model and is a widely discussed topic in the literature.   The solutions representing the tensor-mode excitations in the presence of a brane are examined in~\cite{Cui:2020fiz,Xu:2022xxd} with a focus on the stability problem.  It is proven that the extra-dimensional part of the metric is perturbatively stable provided that conditions \eqref{ghost} are valid.
The stabilization of extra space as a pure gravitational effect has been studied in~\cite{2002PhRvD..66d4014G}.
Analytical solutions allow for prompt study of the stability~\cite{Xu:2014jda,Xu:2022xxd,Liu:2009dt} of one-brane configurations in 5D space--time, and it is of interest to apply these results to the stability question in the presence of the two branes discussed here.

The stabilization mechanism for the radion in two-brane  models was proposed in~\cite{Csaki:1999mp}. {The numerical stability of extra dimensions was discussed in our paper~\cite{Nikulin:2020nub}. In the paper~\cite{Petriakova:2023klf}}, we used an alternative method---transforming the initial equation into one resembling the Schrödinger equation.

The introduction of the warp factor in a D-dim metric~\cite{Oda_2000} facilitates analysis, especially in the fermion sector, as outlined in~\cite{Wan:2023usr} which provides a detailed analysis of this subject in even-dimensional space--time. Fermions in other types of extra metrics are studied in~\cite{Gogberashvili_2007,Dantas:2015dca}.

Two-brane models are often based on the well-known 5D Randall--Sundrum model, describing two 3D thin branes located at two fixed points of the orbifold. The model has a fixed distance between the branes, which is one of its disadvantages.  A substantial amount of references can be found in~\cite{Olechowski:2024wcf}. The formation of multiple thick branes is considered in, e.g.,~\cite{Bronnikov:2007kw}.

In this study, we discuss a new class of branes with properties depending on initial conditions that form under the horizon at high energies. There are no a priori assumptions on the existence of branes. We consider purely gravitational action with the simplest extension to gravity with higher derivatives; see \eqref{SfR} and \eqref{fR}.
Our analysis demonstrates that one-brane metrics represent exceptions rather than the norm, and we are particularly interested in the set of metrics that describe two branes, located at a finite distance from each other. In this framework, the distance between the branes depends on the  initial conditions.

Here, we analyze two-brane configurations in the framework of the $f(R)$ gravity acting in 6D space--time.
It is assumed that the observers are located on only one of the branes (brane-1). Therefore, the initial conditions should be adjusted  so that the effective physical constants coincide with the experimental ones on brane-1. The problem can be solved by taking into account the random distributions of the matter fields over the extra dimensions due to the quantum fluctuations at the high energies~\cite{2021arXiv210908373R}. The particle masses and their coupling constants are arbitrary on the second brane (brane-2).

The paper is organized as follows.
Section \ref{smetric} is devoted to a comprehensive discussion of metric choice.
In Section~\ref{branes},
we present the results of numerical simulations leading to brane formations in the framework of pure $f(R)$ gravity.
Section~\ref{matter} is dedicated to the exploration of the different behavior of matter fields on branes. The conclusions
of our study are summarized in Section~\ref{Conclusion}.

\section{The Metric Choice}\label{smetric}

In this paper, we study the static extra metric, so we have to act on scales where the wavelengths of 4D fluctuations are much larger than the extra space size.
The D-dim Plank mass $m_D$ determines the highest possible energy scale, and we assume that $m_D\sim 10^{17}\div 10^{18}$~GeV. The scale of extra-dimensional space, which is of the order of the first KK mode, is $l_D\sim 10^{2}m_D^{-1}$, see Figure~\ref{f1}. At the same time, the characteristic wavelength at the de Sitter stage is of the order of the inverse Hubble parameter $H^{-1}$.  Therefore,  estimation $H\equiv H_D\ll l_D^{-1}\sim 10^{15}\div 10^{16}$~GeV provides us with the energy scales where the extra dimensional metric can be considered as static. Here, we keep in mind that the wavelengths larger than the scale of the extra dimensions cannot disturb the extra metric.

The general picture is as follows. Our universe is created at the D-dim Planck scale $m_D$ due to strong quantum fluctuations of the metric and fields. Then, the universe moves slowly down to the scale $H_D$ while the extra metric is heavily influenced by the quantum fluctuations. Simultaneously, the 4D de Sitter space is divided into the causally disconnected spatial domains,  each of which has a specific extra metric. The extra metric is frozen at the energies below $H_D$ and remains constant thereafter.
Some processes continue to produce the 4-dim field fluctuations that are currently observed. However, their energy is too small to disturb the extra-dimensional metric. In other terms, their wavelengths are much larger than the extra-dimensional size, and we will {use the approximation}
\begin{equation}\label{der0}
\partial_{\mu} \cdot\ll\partial_a \cdot, \quad \mu=1,2,3,4,\, a=5,6
\end{equation}
in the range $H<H_D\sim 10^{14}\div 10^{15}$~GeV to study the static classical distributions over the extra dimensions.

Another factor is important in the analysis, although not essential. In the subsequent discussion, the assumption $H \simeq \text{const}$ (more accurately, $\epsilon=\dot{H}/H^2 \ll 1$) is often used, and evaluating its precision is essential. There are two energy scales where this approximation is accurate. The first  is above the threshold $H>H_I\sim 10^{13}$~GeV, marking the end of inflation. The second, the lowest energy range, is given by $H \simeq \sqrt{\Lambda/3}\simeq 0$, where $\Lambda$ is the current cosmological constant. Both of these limits do not contradict the interval $H\simeq 10^{14}\div 10^{15}$~GeV obtained above. The analysis above indicates that metric \eqref{metric} holds in the energy interval $(H_I, H_D)$ at high energies. The second appropriate interval lies at low energies, $0\leq H\ll H_I$.


As a result, we will consider the 4D conformal de Sitter metric with a 2D factor space in the form
\begin{align}\label{metric}
ds^{2} = &g_{MN} dX^M dX^N
=e^{2\gamma(u)}\Bigl(dt^2 - e^{2Ht}\delta_{ij}dx^i dx^j\Bigr)
\nonumber \\ &
- du^2 - r^2(u)\,d\Omega_{n-1}^2 \, , \quad i,j=\overline{1,3} \, ,
\end{align}
with the radial extra coordinate $X^4 \equiv u$ and $X^{\mu}\equiv x^{\mu}, \mu=0,1,2,3.$ The angular coordinate $X^5=\theta$ varies in the finite region. We consider 6D space so that $D=4+n,\, n=2$ and the set of extra dimensional coordinates is $y=\{X^4,X^5\}=\{u,\theta\}$. The brane position in the radial extra coordinate is usually postulated~\cite{Dantas:2015dca,2000PhRvL..85..240G, Oda_2000}. { By brane, we mean the subspace of the $4+{n}$-dimensional space--time under consideration in which the material fields are concentrated. } In our study, the position of the brane $u_*$ is determined by solving the equations of $f(R)$ gravity. They are formed spontaneously with the form determined by random initial conditions. The origin of the latter is closely related to the metric fluctuations during inflation.

Sometimes, relations between the metric functions $e^{\gamma(u_*)}$ and $r(u_*)$ are postulated to avoid possible singularities ~\cite{Dantas:2015dca}.
In this paper, we do not impose such restrictions, assuming that the formed singularities are weak enough to be integrated.
This subject has been discussed earlier~\cite{Rubin:2015pqa,Petriakova:2023klf,Bronnikov:2023lej}; see also the text around expression \eqref{f*R} of this paper.  Such sharp peaks (the cusps) have been also discussed in a variety of papers, mostly in 5D models; see, e.g.,~\cite{Guo:2024izl}.

Consider $f(R)$ gravity in a $\D = 4 + n$-dimensional manifold $m_\D$:
\begin{eqnarray}\label{SfR}
S = \frac{m_{\D}^{\D-2}}{2}  \int_{M_\D}  d^{\D} X \sqrt{|g_{\D}|} \,   f(R)
\, ,
\end{eqnarray}
 where $g_{\D} \equiv \det g_{\M\N}$, $\M,\N =\overline{1,\D}$,
 the $n$-dimensional manifold $M_n$ is assumed to be closed, $f(R)$ is a function of the
 D-dimensional Ricci scalar $R$, and $m_\D$ is the $\D$-dimensional Planck mass. Below, we
 will work in the units $m_D=1$.
Variation in the action \eqref{SfR} with respect to the metric $g^{\M\N}_\D$ leads to the known equations
\begin{align}         \label{eqMgravity}
    &-\frac{1}{2}{f}(R)\delta_{\N}^{\M} + \Bigl(R_{\N}^{\M} +\nabla^{\M}\nabla_{\N}
        - \delta_{\N}^{\M} \Box_{\D} \Bigr) {f}_R  = 0,
\end{align}
  with $f_R = {df(R)}/{dR}$, $\Box_{\D}= \nabla^{\M} \nabla_{\M}$. We use the conventions for the curvature tensor $R_{\ \M\N\K}^\LL =\partial_\K\Gamma_{\M\N}^\LL-\partial_\N \Gamma_{\M\K}^\LL +\Gamma_{\CC\K}^\LL\Gamma_{\N\M}^\CC-\Gamma_{\CC\N}^\LL \Gamma_{\M\K}^\CC$ and the Ricci tensor $R_{\M\N}=R^\K_{\ \M\K\N}$.

We assume that the Hubble parameter $H(t)$ is a slowly varying function of time, consistent with the inflationary paradigm. Above, we justified the approximation \linebreak $H(t)\sim H \simeq const$, which significantly simplifies the analysis. 
We also set $\partial_{\mu}\cdot =0$ in all the following equations, according to \eqref{der0}.   Furthermore, we assume the s-wave approximation, omitting the dependence on the extra angle coordinate $\theta$.
Consequently, the extra metric and the field distributions depend only on the internal radial coordinate $u$ at the de Sitter stage.

As a result, the de Sitter stage permits us to exclude the 4-dim coordinates dependence from the equations at high energies. The only trace of time remains in the Hubble parameter $H(t)$, which varies extremely slowly with time.

The system \eqref{eqMgravity} is the system of differential equations with respect to the argument $u$, according to the discussion above. The solutions form a continuum set of functions depending on additional conditions $\gamma(u_0),\, d\gamma(u)/du\ |_{u=u_0}, ...$ at an arbitrary chosen point $u_0$. In the next section, we discuss the solutions of system \eqref{eqMgravity}, the explicit form of which is given in our previous papers~\cite{Petriakova:2023klf,Bronnikov:2023lej}.

\section{Brane Configurations}\label{branes}

In this section, we show that the metrics obtained as a solution to the classical equations relate to two sorts of manifolds with different topology. The first type of manifold represents the closed extra space endowed with two branes. The second type of the extra space represents { the open space with two branes}. We do not choose the topology a priori. The explicit form of the numerical solution to system
\eqref{eqMgravity} has also been intensively discussed in our previous papers~\cite{Petriakova:2023klf,Bronnikov:2023lej}.
Here, we study the behavior of matter in the vicinity of branes. We define a brane as a 3D hypersurface characterized by a cusp in the Ricci scalar. { The geometric configuration of the subspace thus defined depends on the selected coordinate system. The coordinate choice used in \eqref{metric} provides the most straightforward representation of the relative positioning of the branes.
}

Here, we restrict ourselves to pure $f(R)$-gravity in its simplest extension of the Einstein's gravity in the form
\begin{equation}\label{fR}
  f(R)=aR^2 +R +c.
\end{equation}
{The matter}
 fields are considered as test fields in the subsequent sections. We also assume that the ghost- and the tachyon-free conditions
\begin{equation}\label{ghost}
  df(R)/dR > 0  ;\quad d^2 f(R)/dR^2>0
\end{equation}
hold.

The general solution of the system \eqref{eqMgravity} is a hard problem. Nevertheless, under the conditions discussed above, the extra-dimensional metric depends only on the radial coordinate $u$ of the extra space. This greatly simplifies the analysis so that non-trivial numerical results can be obtained. We refer the reader to the system of differential {Equations}
 (6)--(11) written in~\cite{Bronnikov:2023lej}. The brane position depends on a certain solution which is characterized by additional conditions: specific values of metric functions and their derivatives at the point $u=u_0$. We choose $u_0=0$ without loss of generality.

\begin{figure}[H]
\centering
\includegraphics[width=0.25\textwidth]{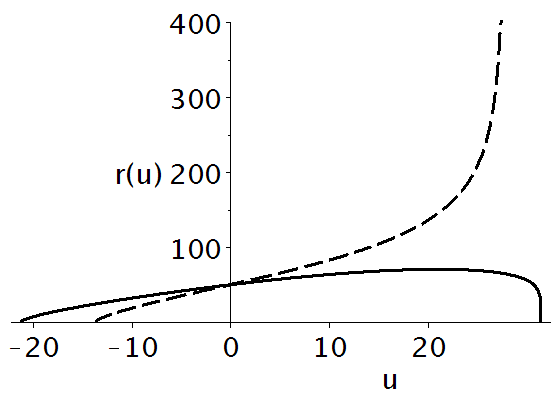} \  \includegraphics[width=0.3\textwidth]{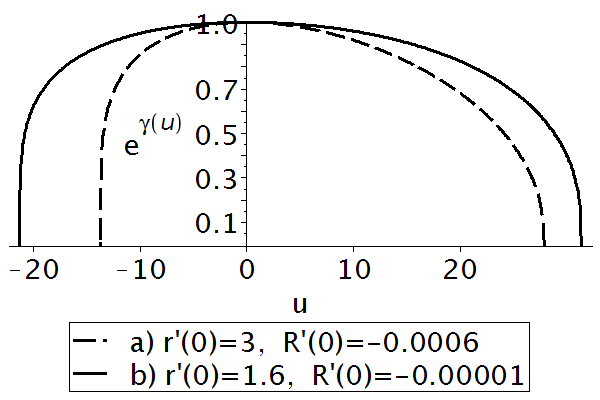} \ 
\includegraphics[width=0.25\textwidth]{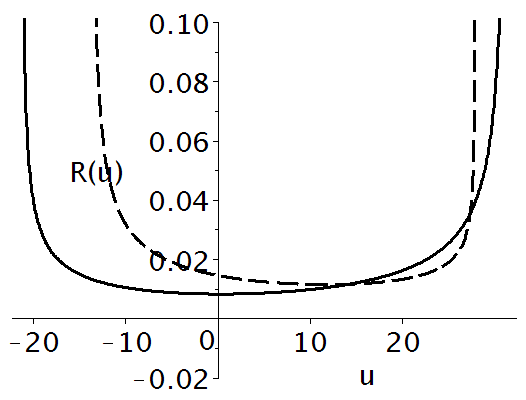}
\caption{ \footnotesize{
The numerical solution of Equation~\rf{eqMgravity} in units $m_D=1$,  for the following parameters: $n=2$,  $H=0$  and the boundary conditions  $r(0) =50$, $\gamma(0)=0, \gamma'(0) =0$, 
(a) $r'(0) = 3$,  $ R'(0) = -0.0006, R(0) \simeq  0.01451$, $ u_{1} \simeq -13.671, u_{2}  \simeq 27.865 $; 
(b) $r'(0) = 1.6$, $ R'(0) = -10^{-5},$ $  R(0) \simeq  0.00824$, $u_{1} \simeq -21.244, u_{2}  \simeq 31.374$.
}}
\label{f1}
\end{figure}

The results for numerical simulations are shown in Figure~\ref{f1}. Note that all results are obtained for fixed Lagrangian parameters
$$a=300, c=0.02$$
and different  additional conditions. The solid curve on the left panel describes the behavior of the extra-space radius $r(u)$. It crosses the abscissa at two points $u_1$ and $u_2$, which means that the closed extra-space is formed.

The right panel in the figure describes the curvature of the extra space and indicates that singularities  or cusps can exist at points $u_1$ and $u_2$. They are quite weak, so that the integration over them does not lead to infinities~\cite{Bronnikov:2023lej}. Such cusps can be suppressed by imposing artificial relations between metric functions~\cite{Dantas:2015dca}. Another way to avoid the singularities is to choose a more complicated function $f(R)$, like
\begin{equation}\label{f*R}
    f_*(R)=aR^2\e^{-\epsilon R^2} +R +c,\, 0<\epsilon \ll 1.
\end{equation}
Suppose that at some point $R\to \infty$. Then, $f_*(R)=R +c$ with the well-known solution in the form of the de Sitter metric. (Remember that the matter is still absent). Thus, the curvature $R\to {const}$ contradicts the initial assumption. This means that the limit $R\to \infty$ is not realized. The exact solution to the system \eqref{eqMgravity} with the function $f_*$ is a hard problem. At the same time, the multiplier $\e^{-\epsilon R^2}\simeq 1,\, (\epsilon \ll 1)$ with an arbitrary good accuracy near singularity if the parameter $\epsilon$ is small enough. Hence, we continue our numerical calculations with the quadratic function $f(R)$, assuming that the possible curvature singularity is replaced by a sharp maximum.

\begin{figure}[H]
\bigskip
\centering
\includegraphics[width=0.3\textwidth]{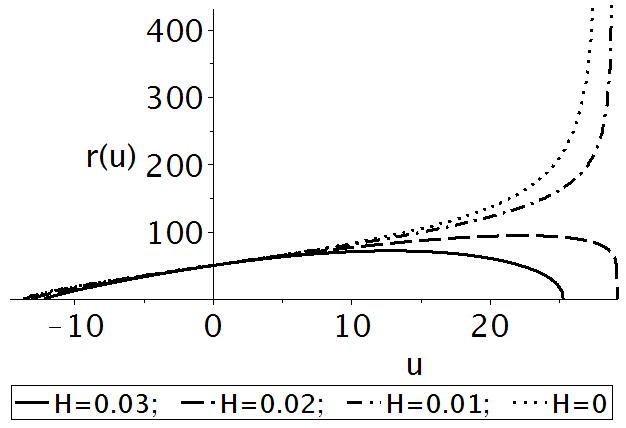} \
\includegraphics[width=0.25\textwidth]{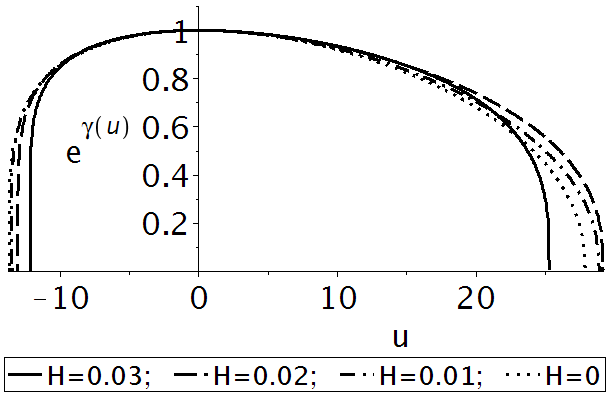}\
\includegraphics[width=0.25\textwidth]{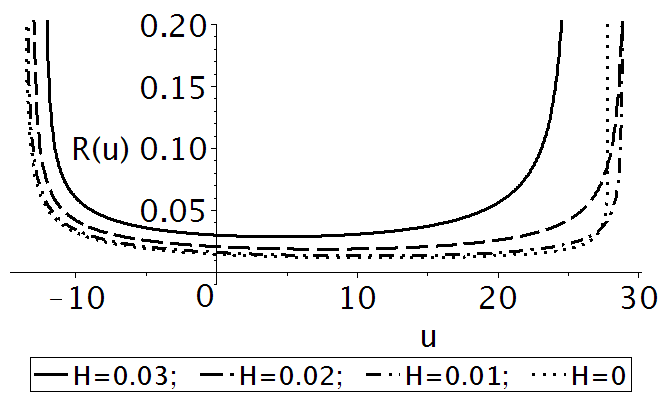}
\caption{\footnotesize {The numerical
 solution of Equation~\rf{eqMgravity},  for the following parameters $n=2$, $f(R) =300 R^2 +R +0.02$} and the boundary conditions  $r(0) =50$, $\gamma(0)=0$,
$r'(0) = 3$, $\gamma'(0) = 0, R'(0) = -6\cdot 10^{-4},$
$\text{(a)}
  H=0.03, R(0) \simeq  0.02986, \text{(b)} \ H=0.02, R(0) \simeq  0.02060, \text{(c)}\ H=0.01, R(0) \simeq  0.01590, \text{(d)} \ H=0, R(0) \simeq  0.01451$.  }
\label{fnonc}
\end{figure}

The dashed lines in the figure relate to another metric which is topologicaly different from the first one. The nontrivial effect of
transition from one topology to another is illustrated in Figure~\ref{fnonc}. The lines describe the extra metrics' behavior as the solutions to the classical equations with the same Lagrangian parameters and additional conditions. The only difference is the Hubble parameter value, which slowly decreases together with the energy scale. Evidently, there exists a  value of the Hubble parameter at which the topology changes. We postpone the discussions on this phenomena to a later time. The stability of such branes is discussed in ~\cite{Cui:2020fiz}.

In the next section, we study the way to concentrate a matter in 3D space within the local extra-space region, which would mean the formation of a brane. In our previous paper~\cite{Rubin:2015pqa}, we mentioned the effect of the scalar field concentration near the critical points where the curvature has a sharp peak. Here, a more thorough analysis is performed.

\section{Matter Localization on Brane}\label{matter}
Here, we study the matter distribution along the radial coordinate $u$ in the vicinity of a brane. We focus on the point $u=u_1$, which is characterized by the sharp peak of curvature. The geodesic motion of the point-like mass as a first test of a possible way to trap particles in the local region around $u_1$ has been discussed in~\cite{Bronnikov:2023lej}, {Appendix A.}
 It was shown there that at least non-relativistic massive particles are trapped by branes.
Indeed, their motion is governed by the equation
\begin{equation}
\ddot{u}\simeq -2\text{e}^{2\gamma}\gamma'.
\end{equation}
Therefore, the particles are accelerated to a brane as it is seen from the behavior of warp factor $\gamma(u)$ on Figure~\ref{f1}, the second panel.
%


Our goal is to find the distribution of fields (scalars, fermions, vectors) over the radial extra coordinate $u$ and to study them in detail.
We will seek for the field solution in the form
\bear \label{decomp}
o(x,y)&=&O(x)Y(y)+\sum_{k=0} o_k(x)Y_k(y)
\simeq O(x)Y(y)
\ear
The orthonormal functions satisfy the equation
\begin{equation}
\Box_n  Y_k(y)   =\lambda_k Y_k(y)
\end{equation}
where $\Box$ is the D Alembert operator acting in $n$-dimensional extra space. The functions $o, o_k$ and $O$ are assumed to be endowed with equal group indices depending on the group representation, while the scalar function $Y$ is responsible for the distribution over the extra dimensions.
We consider the wavelengths much larger than the scale of the extra dimensions. Consequently, the summation accounting for field variations across the extra dimensions may be disregarded.

The task is made considerably easier if we keep formula \eqref{der0} in mind, so that $O(x)\simeq \text{const}$
with great accuracy. This helps to fix the extra-dimensional distribution $Y(u)$  of the fields as a solution of the system \eqref{eqMgravity}. The knowledge of the extra-space distributions  allows us to substitute decomposition \eqref{decomp} into the initial action and to integrate out extra coordinates, thus reducing the D-dim action to the observable 4-dim one.

\subsection{Scalar Field}\label{scalar}
The aim of this subsection is to find the scalar-field distribution over the extra space, which is formed and stabilised at very high energies.
The conjecture to be proved is that particles (field fluctuations) can be concentrated on the branes.
The classical equation for the free massive field has the standard form
\begin{equation}
    \Box_{\D} \, \varphi + W'_{\varphi}(\varphi) =0;\quad W(\varphi)=\frac {m^2}{2} \varphi^2 .
\end{equation}
Let us find the scalar-field distribution over extra dimensions to integrate them out from the initial D-dim action. To this end, we will seek solutions in the form of \eqref{decomp}.
The method of variables separation leads to the system
\begin{eqnarray}
&&  \Box_4 \Phi(x)+\mu^2 \Phi(x)=0 \label{eqfi} \\
&&-\left[\partial^2_u+\left(4\gamma' +(n-1)\dfrac{r'}{r}\right)\partial_u\right] Y(u)
+(m^2 -\mu^2)Y(u) = 0 ,\label{extraeq}
\end{eqnarray}
where $\varphi(x,u)=\Phi(x)Y(u)$ and $\mu$ is an arbitrary parameter that can be fixed by observations. The parameter $\mu$ varies in the range $0\leq\mu\leq m_D$ and is known as the Kaluza--Klein tower of masses if $\gamma = 0,\, r(u)=r_{KK}=const$ and equals $\mu=k/r_{KK},\, k=0,1,2...$.
Note that the extra metric considered in this paper is not periodic in the coordinate $u$.

%

A new possibility opens up when the extra-space dimensionality $n>1$. There is the continuum set of static solutions to the Equation \eqref{extraeq}, because solutions do not satisfy the periodicity conditions. These solutions together with the metric functions $\gamma(u),r(u)$ depend on the Hubble parameter $H$, which decreases slowly. At the same time, if we keep in mind the inflationary paradigm, the Hubble parameter $H$ varies quickly if $\mu >H$. Therefore, we have to choose a solution with small values of the parameter $\mu\ll H\sim m_D\equiv 1$. In this case, we can study solutions to the Equation \eqref{eqfi}, assuming that the 4-dim space is endowed with the de Sitter metric with a good approximation. 

\begin{figure}[H]
\centering
\includegraphics[width=0.35\textwidth]{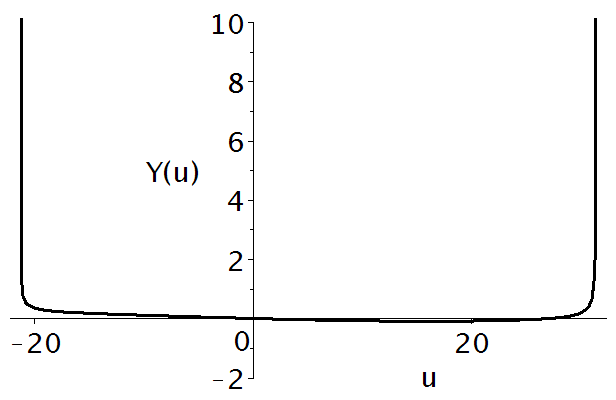}
\caption{ \footnotesize The scalar-field distribution along the radial coordinate $u$ for the gravitational background described in Figure~\ref{f1}b, $m=0.1$  and the boundary conditions $Y(0)=0, Y'(0)=-0.01$. }
\label{Yu}
\end{figure}

\begin{figure}[H]
\centering
\includegraphics[width=0.3\textwidth]{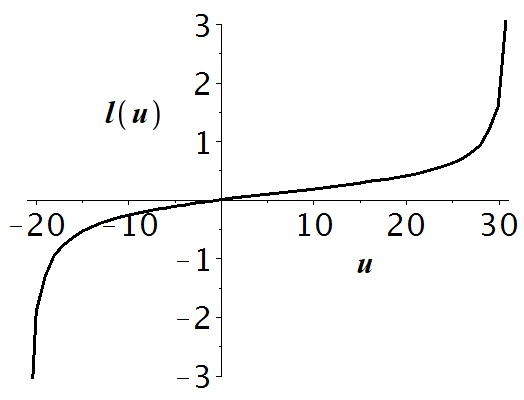} \quad \includegraphics[width=0.4\textwidth]{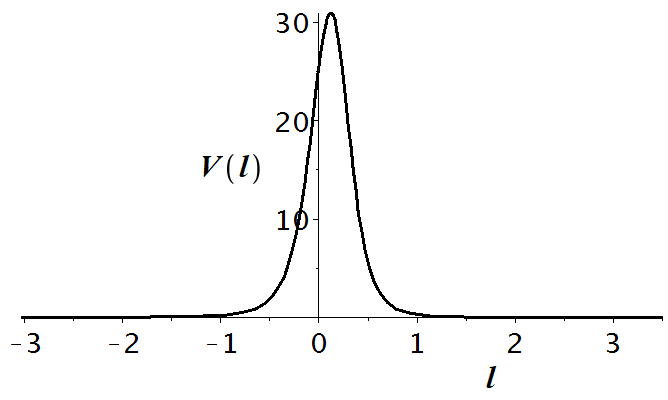}
\caption{ \footnotesize {$l(u)$}
 and $V(l)$   for the gravitational background described in Figure~\ref{f1}b. }
\label{Vsc}
\end{figure}

A typical numerical solution of Equation \eqref{extraeq} is shown in Figure~\ref{Yu}, where we neglect $\mu$ compared to $m\sim m_D$. One can see that the scalar field is indeed concentrated near the points $u_{1,2}$, leaving desert with $Y\simeq 0$ in between. Let us estimate the effect analytically by reducing Equation \eqref{extraeq} to the stationary Schrodinger-like equation.
Moving on to a new variable $u \rightarrow l $
\begin{equation}\label{dldu}
\frac{d l}{d u} = \frac{1}{ r^{n-1} \e^{4 \gamma}  }\, ,
\end{equation}
Equation \eqref{extraeq} can be rewritten as follows
\begin{equation}\label{shred}
\frac{d^2 Y}{d l^2} -V(l) Y =0, \quad {V(l)=  m^2\,   r^{2n-2}\, \e^{8 \gamma}}\, , 
\end{equation}
where the typical forms of $l(u)$ and $V(l)$ is shown in Figure \ref{Vsc}.
If $V\gg 1$, we may use the quasi-classical approach to solve this equation. Let us insert the small parameter $\epsilon$ so that $V=\tilde{V}/\epsilon$ and change variables $Y=\e^{S/\epsilon}$.  Substitution in Equation \eqref{shred} gives two solutions
\bear \label{scaldist}
Y_{1,2} (l) &=& Y_{\pm}\exp\left(\pm \int^l \sqrt{V(l')}dl'\right)
= Y_{\pm}{\exp\left(\pm m\int^lr(l')^{n-1} \e^{4\gamma(l')}dl' \right)},
\ear
where we have neglected terms $\propto 1/\epsilon$ in comparison to $1/\epsilon^2$. Coming back to the coordinate $u$, see \eqref{dldu}, we obtain
two functions describing the scalar field in both sides of the potential maximum
\begin{eqnarray}
&&Y_{1}(u) =Y_{-} \exp\left[ m(u_1-u)\right], \\
&&Y_{2}(u) =Y_{+} \exp\left[ m(u-u_2)\right].
\end{eqnarray}\label{Sdistr}
\hspace*{-1mm}The quasi-classical approximation is violated in the vicinity of the brane, so the obtained result describes the field behavior only between the branes. Therefore, the 4D fluctuations near each brane interact extremely weakly with each other, due to the smallness of the overlapping integral
\bear \label{scaldist2}
\int du Y_1(u) Y_2(u) &=& Y_ -Y_+  \int_{u_1}^{u_2} du \exp{\big[m(u_1-u)\big]}
\exp{\big[m(u-u_2))\big]}
\nn &&
\sim  \exp\big\{m(u_1-u_2)\big\}\ll 1.
\ear
The influence of the scalar field concentrated on brane-2 on the processes on brane-1 is negligibly small, and vice versa. An important conclusion is that we can study the fields on each brane independently. As we will see below, the low-energy physical parameters depend on the extra metric and thus can be different on both branes.

\subsection{Fermions}\label{fermions}
Here, we study the fermion field distribution along the radial extra-dimensional coordinate $u$.
The general form of the free fermion action in D dimensions can be found, for example, in~\cite{Liu:2017gcn}.

\begin{equation}\label{Sf}
    S_f=\int d^D z\sqrt{|g|}\left[ \bar{\Psi} i\Gamma^M D_M \Psi \right] .
\end{equation}
The mass term is assumed to be generated by the interaction with the Higgs field. A more thorough model in even dimensions is outlined in~\cite{Wan:2023usr} where conditions for decoupling the left and right chiralities of spinors are studied.

The particular form of the curved gamma matrices $\Gamma^M$, the spin connection $\omega_A$, and the covariant derivatives can be found in~\cite{Oda_2000}.
In this subsection, we follow that paper and refer the reader to the details there. A useful discussion is also performed in
\cite{Appelquist:2000nn,Randjbar-Daemi:2000lem}.

A 6D spinor
\begin{equation}\label{2bispinor}
    \Psi =\left(\begin{array}{c}
  \xi_1 \\
  \xi_2
\end{array}\right)
\end{equation}
 acting in six dimensions is equivalent to a pair of 4-dimensional Dirac spinors $\xi_1$ and $\xi_2$. As shown in~\cite{Gogberashvili_2007}, both 4D spinors, $\xi_1$ and $\xi_2$, are equal in the s-wave approximation used in our approach. The more complicated situation with the contribution of p-waves ($l\neq 0$) is also discussed there. In this paper, we are interested in the spinor distribution over the extra dimensions. We will derive a subsequent representation of the fermion field in the form \eqref{decomp} discussed in the previous section. Namely,
\begin{equation}\label{fermion}
\Psi(x,y)=\psi(x)Y_f(y)+\sum_{k=1} \psi_k(x)Y_k(y) \simeq \psi(x)Y_f(y).
\end{equation}
In this context, $\psi_k(x)$ denotes massive Kaluza--Klein (KK) modes, which are excluded by concentrating on the zero-mode approximation. The spinor $\psi$ is characterized by both spinor and group indices and functions within a six-dimensional framework. This study is focused on the distribution of fields across the extra-dimensional coordinates rather than their reduction to four-dimensional physics. Consequently, our attention is directed toward the form of the function $Y(y)$ exclusively.

The explicit form of the scalar function $Y_f(u)$ is needed to study whether it is really trapped on the brane.
To this end, consider the equation of motion
\begin{eqnarray}\label{ef1}
&&0={i\Gamma^M D_M \Psi}=\Gamma^{{M}}(\partial_M + \omega_M)\Psi \nonumber\\
&&=[\Gamma^{\mu}(\partial_{\mu} + \omega_{\mu})+\Gamma^{u}(\partial_{u} + \omega_{u})+\Gamma^{a}(\partial_{a} + \omega_{a})]\Psi \nonumber \\
&&=\Gamma^{\mu}\partial_{\mu}\Psi(y|x)+\hat{O}(y|x)\Psi(y|x).
\end{eqnarray}
The matrix form of the operator $\hat{O}$
\begin{eqnarray} \label{O}
\hat{O}(y|H,x)&\equiv& \Gamma^{0}\omega_{0}+\Gamma^{i}\omega_{i} +\Gamma^{u}(\partial_{u} + \omega_{u})
+\Gamma^{a}(\partial_{a} + \omega_{a})
\end{eqnarray}
depends on the dimensionality of the space.

Keep in mind that the 4D derivatives of fields are small compared to the derivatives in the extra space; see \eqref{der0}. Therefore,
$\partial_{\mu}\Psi \cong 0$
and we obtain the equation
\begin{equation}\label{psiu}
\hat{O}(u|H)Y_f(u)\cong 0
\end{equation}
for the field distribution along the extra coordinate $u$.

\begin{figure}[H]
\centering
\includegraphics[width=0.35\textwidth]{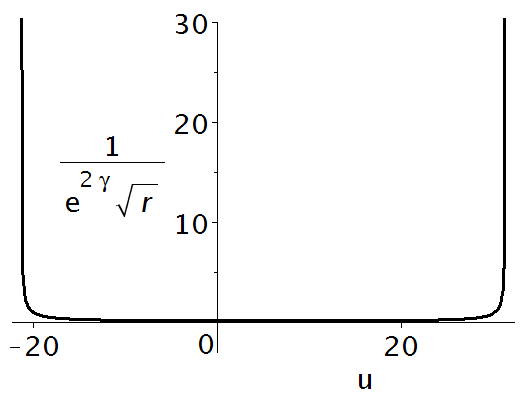}
\caption{ \footnotesize The fermion distribution along the radial coordinate $u$ for the gravitational background described in Figure~\ref{f1}b. }
\label{Y}
\end{figure}

The distribution $Y_f(u)$ can be found by solving the Equation \eqref{psiu}, which is transformed into the ordinary differential equation
\begin{equation} \label{eqY}
\Big[\partial_u + \frac{r'(u)}{2r(u)} + 2 \gamma'(u)\Big]Y(u)=0.
\end{equation}
The solution in the analytical form
\begin{equation}\label{Yfermi}
Y(u)=c_2r(u)^{-1/2}e^{-2\gamma(u)}
\end{equation}
was obtained in~\cite{Oda_2000}, {Section 4.3.}
The metric functions $e^{\gamma(u)}, r(u)$ tend to zero on branes; see Figure~\ref{Y}. This means that the solution \eqref{Yfermi} has a sharp peak on branes and, hence, the fermions are located there.
There is also a region between the two branes where the field distribution and its derivatives are close to zero. Therefore, the interference between fields located on different branes is negligible, as has been discussed in Section \ref{scalar}. We have obtained two independent worlds located on the branes. The fields on different branes do not affect each other. Note that if the extra metrics on the two branes are different, then the effective parameters of the fields are also different. Keeping this in mind, we can limit ourselves by fixing the 4-dim low energy physics to only one brane.

\subsection{Electromagnetic Field}

The metric \eqref{metric} contains the warp factor $e^{2\gamma(u)}$, which is responsible for the gauge-field localization on string-like defects~\cite{Oda_2000}. It holds even if a gauge field does not depend on the internal coordinates, which are $u$ in our case. In this subsection, we show that the unique distribution of the Abelian gauge field exists and does not depend on the extra coordinates.

The action of an Abelian gauge field has the standard form, see e.g.,~\cite{Dubovsky:2000av}
\bear
 && S_{gauge}=-\frac {\text{g}_D}4\int d^D X \sqrt{|g_{\D}|}g^{PM}g^{QN}  F_{PQ}F_{MN},
 \nn &&
 F_{QN}=\partial_Q A_N - \partial_P A_Q
\ear
Here, $\text{g}_D$ is the D-dimensional coupling constant.
Field equations
\begin{equation} \label{efe}
\partial_P(\sqrt{|g_D|}g^{PQ}g^{MN}F_{QN}) =0
\end{equation}
are used to find the field distribution along the extra coordinate $u$.
Keeping in mind the expression \eqref{decomp} in the form
\beq
A^N(x,u)=a^N (x)Y_{em}(u)
\eeq
and the condition $\partial_{\mu}\cdot =0$ used above, Equation \eqref{efe} can be rewritten as follows ($n=2$)
\begin{eqnarray}\label{mu}
&&M=\mu :\quad a^\mu \left[ Y_{em}'' +\left( 2\gamma' +\frac{r'}{r} \right)Y_{em}' \right]=0, \nn
&& M=\theta:\quad a^{\theta} \left[ Y_{em}'' +\left( 4\gamma' -\frac{r'}{r} \right)Y_{em}' \right]=0.
\end{eqnarray}
Here, only nontrivial equations are written. Evidently, $Y_{em}(u)=C_{em}=const$ is the unique solution to system \eqref{mu} if $a^{\mu},\, a^{\theta}\neq 0$.
Hence, there exists the unique classical solution
\beq
A^N(x,u)=C_{em}a^N(x)
\eeq
and it does not depend on the internal coordinates.  $C_{em}$ is an arbitrary constant. Note that the gauge condition choice is not necessary to prove this statement.

Finally, the effective low-energy action is as follows
\begin{eqnarray}
S_{gauge}&=&-\frac {\text{g}_D}4\int d^D X \sqrt{|g_{\D}|}g^{\mu\nu}g^{\alpha\beta} F_{\mu\alpha}F_{\nu\beta} \nn
&& = -\frac {{\text{g}_4}}4\int d^4 x \sqrt{|g_{(4)}|}g_{(4)}^{\mu\nu}g_{(4)}^{\alpha\beta}  f_{\mu\alpha}f_{\nu\beta},  \end{eqnarray}
where
\begin{equation}f_{\mu\nu}\equiv\partial_{\mu} a_{\nu} - \partial_{\nu} a_{\mu},
\end{equation}
and
\begin{equation}
{\text{g}_4}\equiv 2\pi C_{em}^2\int_{u_1}^{u_2} du r(u)\e^{4\gamma(u)}
\end{equation}
is the effective 4-dim coupling constant. It is also assumed that the vector projections $a^u$ and $a^{\theta}$ are not perturbed by 4-dim fluctuations at low energies, and hence do not depend on $x.$. One can conclude that the photons are disposed between the branes.

The exploration of non-Abelian gauge fields presents a significantly more complex challenge. Nonetheless, it is simplified to the previously addressed Abelian case when the field amplitudes are small. In this case, the nonlinear terms of the gauge-field strength
\begin{equation}
    F^a_{MN}\equiv \partial_M A^a_N - \partial_N A^a_M + f^{abc} A^b_M \, A^c_N \, , \qquad a = 1,2,3
\end{equation}
can be neglected. Here, the constants $f^{abc}$ are the structure constants of the Lie algebra of the generators of the gauge group.

\subsection{Gravitational Fluctuations}\label{grav_}

In this subsection, we follow the paper~\cite{Cui:2020fiz}, which discusses the linear stability of space--time under gravitational perturbations. The metric is chosen in the form
\begin{equation}
    ds^2=e^{2\gamma(u)}(\eta_{\mu\nu}+h_{\mu\nu})dx^{\mu}dx^{\nu}-du^2-r(u)^2d\phi^2
\end{equation}
$\eta_{\mu\nu}=diag (1,-1,-1,-1),\quad h_{\mu\nu}\ll \eta_{\mu\nu}$
where the perturbation $h_{\mu\nu}$ represents the tensor mode obeying the  transverse--traceless gauge conditions
$$\eta_{\mu\nu}h^{\mu\nu}=0,\quad \partial_{\mu}h_{\mu}^{\nu}=0$$

After some algebra, see~\cite{Cui:2020fiz}, the equations for the linear tensor perturbations are reduced to a non-trivial equation for the {perturbation}
\begin{eqnarray}\label{graveq}
  \left[e^{-2\gamma} \eta_{\mu\nu}\partial^{\mu}\partial^{\nu}+r^{-2}\partial_{\phi}^2 +\left(4\gamma'+\frac{r'}{r}\right)\partial_u  {+ \partial_u^2} +\frac{\partial_u f_R}{f_R}\partial_u
  \right] {h_{\alpha\beta}}=0
\end{eqnarray}
We are interested in the zero mode
\begin{equation}\label{grav}
    h_{\mu\nu}(x,u) =H^{(0)}_{\mu\nu}(x)Y_0(u)
\end{equation}
keeping in mind decomposition \eqref{decomp}.
Our aim is to find the distributions over the extra dimensions, so we put $H_{k,\mu\nu}(x)=const$  ($x-$derivatives are small, see \eqref{der0}). It is evident that one of the possible solutions to Equation \eqref{graveq} is $Y_0=const$. Therefore, the gravitational field is uniformly distributed between the two branes. Gravitons generated by a matter on the brane-2 can affect a matter on the brane-1 and vice versa.

\subsection{The Higgs Field}

Up to now, we have not included the Higgs field interaction, so that the low-energy fermions remain massless.
In the paper~\cite{Bronnikov:2023lej}, we analyzed the Higgs field properties in D dimensions in the context of the hierarchy problem. Here, we study a new aspect---the localization of the Higgs field on the different branes.

Suppose that the form of the Higgs action at the {Planck scale}
 is the same as at the electroweak scale,
\bear            \label{SH}
&& S_{\HH_P} =
\frac12 \int d^{\D} X \sqrt{|g_{\D}|} \,
\Bigl(\partial^{\M} {H_P}^\dagger \partial_{\M} H_P
+ \nu {H_P}^\dagger H_P - \lambda\bigl({H_P}^\dagger H_P\bigr)^2 \Bigr),
\ear
where the symbol $\dagger$ means Hermitian conjugation, $\nu >0$ and $\lambda  > 0$ are arbitrary numbers, and $H_P$ is a proto-Higgs field.
It is assumed that the parameter $\nu$ is many orders of magnitude larger than the known value of the Higgs vacuum average $v_h$.

Below, we follow the standard procedure used above to integrate out extra coordinates. The classical equations of motion are obtained by varying the action \eqref{SH} with respect to $H_P$, which gives
\begin{equation}            \label{boxHP}
        \square_\D H_P=\nu H_P
        - 2 \lambda  \bigl({H_P}^\dagger H_P\bigr)H_P.
\end{equation}
According to general prescription \eqref{decomp}, the proto-Higgs field can be presented as
\begin{equation}\label{hY}
        H_P = h(x) Y_H(u),
\end{equation}
where $H_P$ and $h(x)$ are two-component columns acting in the fundamental representation of $SU(2)\times U(1)$.
The dimensionality of the proto-Higgs field is $[H_P]=m_D^{(D-2)/2}$, \linebreak $[h]=[h_k]=m_D$,
  $[U]=[Y_k] = m_D^{n/2}$.
Recall that the expression \eqref{der0} holds in the de Sitter metric so that
\begin{equation}            \label{Hv}
h(x) = \frac{1}{\sqrt{2}}
\begin{pmatrix}
0 \\ \rho(x)
\end{pmatrix}
\simeq
\frac{1}{\sqrt{2}}
\begin{pmatrix}
0 \\ 1
\end{pmatrix}.
\end{equation}

\begin{figure}[H]
\centering

\includegraphics[width=0.4\textwidth]{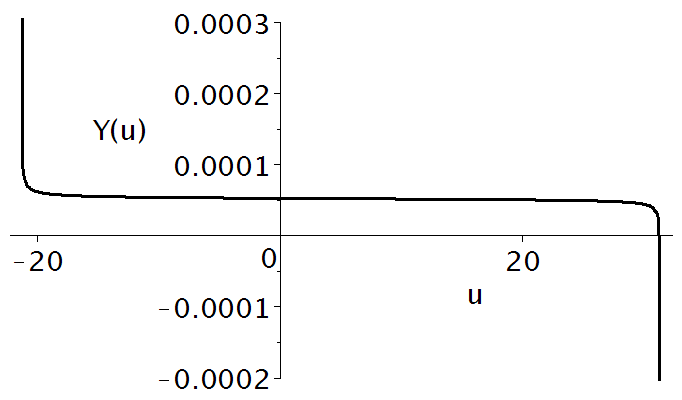}
\caption{\footnotesize
The solution of Equation \eqref{boxUU} on the gravitational background shown in Figure \ref{f1}b. Parameter values:
    $\nu =1 \cdot 10^{-5}$,  $\lambda = 1.9 \cdot 10^{3}$. Additional
    conditions: ${Y}(0) \simeq 0.000052$ (this value corresponds to the stationary state $\sqrt{\nu/\lambda}$ for the $Y_H$ field), ${Y}'(0) = -1 \cdot 10^{-7}$.
    For this solution,  $m_{\Pl} \simeq 333.25\, m_D, \ \lambda_H \simeq 0.13,\ m_H \simeq 10^{-17} m_{\Pl}$. }
\label{YHiggs}
\end{figure}

The approximation \eqref{Hv} transforms Eqaution \eqref{boxHP} in the following way:
\begin{equation}        \label{boxU}
\square_n Y_H(u) =\nu Y_H(u) - \lambda Y_H^3(u),
\end{equation}
or in more details
\begin{equation}        \label{boxUU}
\left[\partial^2_u+\left(4\gamma'+\dfrac{r'}{r}\right)\partial_u\right] Y_H(u) =-\nu Y_H(u) + \lambda\, Y_H^3(u),
\end{equation}
A set of solutions to this equation can be obtained numerically, each for certain additional conditions $Y_H(u_*), Y'_H(u_*)$ at an arbitrary point $u_*$. One of them is represented in Figure~\ref{YHiggs}.
The field is settled mostly in the stationary state with sharp peaks on the branes. The structure of the peaks is different as well as the extra-space metric functions; see Figure~\ref{f1}. The scalar and spinor fields located on the different branes interact with the same Higgs field distributed between the branes.

The extra metric and field distributions over extra dimensions are fixed at high energy at a certain value of the Hubble parameter $H$. The latter slowly approaches the extremely small observable value. This is the reason we perform calculations for $H\simeq 0$; see also the short discussion above formula \eqref{SfR}.

Substitution \eqref{hY} into initial action \eqref{SH} and integration over the extra coordinates $u,\theta$ leads to the 4-dim effective Higgs action
\bear     \label{SHHH}
&&S_{\HH} = \frac{1}{2}\int d^{4} x  \sqrt{|\tilde{g}_{4}|} 
\biggl(\partial_{\mu} H_0^\dagger\partial^{\mu} H_0
+  m_H^2 H_0^\dagger H_0  -  \lambda_H\bigl(H_0^\dagger H_0\bigr)^2 \biggr)\, ,
\\
&& m_H^2 \equiv \frac{m_h^2}{K_h},\  \lambda_H \equiv \frac{\lambda_h}{K_h^2},\   H_0(x)\equiv {h(x)\sqrt{K_h}}.
        \label{mlH}
\ear
The parameters included in the effective 4-dim action \eqref{SHHH}
\begin{eqnarray}           \label{Kh}
K_h = && \mathcal{V}_{\n-1} \int_{u_{1}}^{u_{2}} Y_H^2(u)\,\e^{2\gamma(u)}r^{\n-1}(u)\,du,
 \\
\label{mh}
m^2_h =&& \mathcal{V}_{\n-1} \int_{u_{1}}^{u_{2}} \Bigl(-(\partial_u Y_H)^2 +\nu\, Y_H^2(u)\Bigr)
\e^{4\gamma(u)} r^{\n-1}(u)\, du,
 \\             \label{lh}
\lambda_h = && \mathcal{V}_{\n-1} \int_{u_{1}}^{u_{2}}
\lambda\,Y_H^4(u)\,\e^{4\gamma(u)}
r^{\n-1}(u)\, du, \\
\mathcal{V}_{\n-1} \equiv &&
\int d^{n-1} y \sqrt{|{g}_{n-1}|}
=\dfrac{2\pi^{\tfrac{\n}{2}}}{\,\Gamma\left(\tfrac{\n}{2}\right)},
\end{eqnarray}
depend on $Y_H(u)$ and the extra-space metric. Therefore, the physical parameters slowly vary with time together with the Hubble parameter. The Higgs parameters are measured at the low energies characterised by the Hubble parameter $H\simeq 0$. So, we have to fix all extra-space distributions at $H=0$ as in Figures~\ref{f1} and \ref{YHiggs}.

According to~\cite{Workman:2022ynf},
the experimentally measured parameters, the Higgs boson mass and its vacuum average are
\begin{equation}\label{obs}
        m_{\rm Higgs}=125 \,\text{GeV},\quad v_{\rm Higgs}=246\,\text{GeV}\,.
\end{equation}
They are related to the parameters $m_H$ and
  $\lambda_H$ of the effective Higgs action \eqref{SHHH} as follows:
\begin{equation}            \label{mH}
        m_{H} =m_{\rm Higgs} /\sqrt{2}=88.6\,\text{GeV} \simeq 10^{-17} m_{\Pl},
\end{equation}
where $m_{\Pl}$ is the 4D Planck mass
  and
\begin{equation}\label{lH}
        \lambda_{H}=({m_{H}}/v_{\rm Higgs})^2/2  \simeq 0.13 .
\end{equation}
The appropriate function $Y_H(u)$ is represented in Figure~\ref{YHiggs}.
The effective parameters \eqref{Kh}, \eqref{mh}, \eqref{lh} also depend on the Lagrangian parameters $a, c$ which are fixed from the beginning, and on the distributions $r(u), \gamma(u)$ over the internal coordinate $u$. The latter are randomly created during the fluctuation stage and are specific to each causally separated 3D domain. A Higgs field distribution realized within some of such domains can lead to the observable Lagrangian parameters (see \eqref{SHHH}) and hence resolve the hierarchy problem.


\section{Conclusions}\label{Conclusion}

This research examines the formation of branes in six-dimensional space in the context of pure $f(R)$ gravity. We do not postulate the position and the number of the branes a priori, nor a scalar-field potential that could cause the branes to form. Numerical simulations definitively show that only the two-brane metrics are realized.  The choice of metric \eqref{metric} assumes that the universe is formed at the conformally 4D de Sitter space at high energies ($H > 10^{13}$ {GeV)}
 with a 2D factor space.

It is known that the Hubble parameter decreases with time at an extremely slow rate when chaotic inflation is taken into account, as discussed in reference~\cite{Lindebook}.  Herein, we neglect this time dependence, assuming that $H=const$, and analyze static solutions for different values of the Hubble parameter.

As a consequence of quantum fluctuations, the metrics differ between the infinite number of causally disconnected three-dimensional volumes at high energies. This results in an infinite set of static extra metrics, which have been demonstrated to be the solution to the classical equations. Through numerical analysis, it has been established that there are two types of solutions, \red{namely those with closed and opened extra manifolds.} The infinite set of such solutions is provided by two nonequivalent branes. These branes are settled at some distance from each other. The unique nonsingular solution is discussed in detail in~\cite{Petriakova:2023klf}, {Section 4.}

This article examines the properties of closed extra manifolds with two branes and the manner in which matter fields appears to be attracted to the branes. It is demonstrated that the massive fields are attracted to each brane, leaving the space between them empty. Analytical proof was provided that the electromagnetic field is independent of extra coordinates, thus preserving the observed charge universality. The fermions located on different branes can exchange photons. For example, the positron--electron pair on  brane-1 could annihilate into the positron--electron pair on brane-2.
For the observer located on brane-1, it looks like the positron--electron annihilate into nothing.
This opens the possibility to study such kinds of processes at the LHC collider.

The Higgs field is distributed almost uniformly between branes, acting as a Higgs portal~\cite{Chu:2011be} between matter on different branes. There exists a specific extra metric for which the Higgs vacuum average coincides with the observable one.
The low-energy Lagrangian parameters of the fields are functionals of the extra metric and are specific to each brane.
This leads to the dependence of the physical parameters on the energy scale---in addition to the renormgroup flux.

The research is based on the conceptual model elaborated in~\cite{Petriakova:2023klf,Bronnikov:2023lej}. Necessary elements to be added further are the renormgroup analysis, the standard model in D dimensions, and possible brane interference.

It was shown that the extra metric topology undergoes a transformation at a specific value of the Hubble parameter, as illustrated in Figure~\ref{fnonc}. This non-trivial effect requires a comprehensive investigation in future studies.

\vskip5mm

{\bf Acknowledgments}

{The work of AAP was funded by the development program of the Volga Region Mathematical Center (agreement No. 075-02-2024-1438).
The work of SGR was partially ({Sections} \ref{sec1} and \ref{smetric})
 funded by the Ministry of Science and Higher Education of the Russian Federation, Project "New Phenomena in Particle Physics and the Early Universe" FSWU-2023-0073
and partially \mbox{({Sections} \ref{fermions} and \ref{Conclusion})} by the Kazan Federal University Strategic Academic Leadership Program.}




\end{document}